\documentclass[10pt,aps,pra,amssymb,a4paper,nofootinbib]{revtex4}
\usepackage{amsmath,latexsym,amssymb,amsfonts,amsbsy}
\usepackage{pstricks,pst-all,multido}
\def\vec#1{\boldsymbol{#1}}
\usepackage{times}
\begin{document}
\title{About the stability of the dodecatoplet}
\preprint{LPSC-08-188}
\author{Jean-Marc Richard}
\affiliation{Laboratoire de Physique Subatomique et Cosmologie\\
 Universit\'e Joseph Fourier--CNRS-IN2P3--INPG\\
53, avenue des Martyrs, 38026 Grenoble, France}
\date{Version of \today}

\begin{abstract}\noindent
A new investigation is done of the  possibility of binding the ``dodecatoplet'', a  system of six top quarks and six top antiquarks, $(t^6\bar{t}{}^6)$, using  the Yukawa potential mediated by Higgs exchange.  A simple variational method gives a upper bound close to that recently estimated in a mean-field calculation. It is supplemented by a \emph{lower} bound provided by identities among the Hamiltonians describing the system and its subsystems. 
\end{abstract}
\maketitle
%
%\nocite{*}
\section{Introduction}\label{se:Intro}
The Higgs boson, responsible for generating the masses of fermions in the standard model, couples more strongly to the heavy quarks than to the light ones. The question has been raised whether the attraction mediated by Higgs exchange  could produce new type of bound states 
\cite{Froggatt:2004bh,Froggatt:2008uy,Froggatt:2008hc,Kuchiev:2008fd,Kuchiev:2008gt}.
Higgs exchange corresponds to the Yukawa type of potential
\begin{equation}\label{eq:Yuk}
-\alpha_H \,v(r)=-\alpha_H \frac{\exp(-\mu r)}{r}~.
\end{equation}
 As discussed in the literature \cite{Froggatt:2004bh,Froggatt:2008uy,Froggatt:2008hc}, 
 the coupling of Higgs to the top quark should be about $g_t\sim 1$, with
 $\alpha_H=g_t^2/(4\pi)$. We thus consider $\alpha_H=1/(4\pi)$ as a benchmark value for estimating the spectrum of the potential $v$, but we shall also study how the eigenenergies behave as a function of $\alpha_H$. As for  $\mu$, the Higgs mass, it expected to be is of the order of a hundred or a few hundreds of GeV.

If a bound state of several quarks and antiquarks  occurs due primarily to the above potential, the  precise determination of its mass  in a regime of strong binding should incorporate relativistic effects, strong forces, $W$-exchange inducing $t\bar{t}\leftrightarrow b\bar{b}$ mixing, etc. \cite{Froggatt:2008hc,Kuchiev:2008gt}.
However,  in a regime of weak binding the system remains non-relativistic. 
Thus for a preliminary investigation of the existence of new type of bound states due to Higgs exchanges, it is sufficient to rely on the Hamiltonian~\cite{Froggatt:2004bh,Froggatt:2008uy,Kuchiev:2008fd}
\begin{equation}\label{eq:H}
H_N=\sum _i ^N \frac{\vec{p}_i^2}{2m}-\alpha_H \sum_{i<j} v(r_{ij})~,
\end{equation}
and examine its spectral properties.
If $N$, the number of constituents, does not exceed 6 top quarks and 6 top antiquarks, the colour and spin degrees of freedom can endorse the constraints of antisymmetrisation, and for the orbital variables, the Hamiltonian (\ref{eq:H}) can be considered as acting on effective bosons. 
This is why the attention has been focused on the $(t^6 {\bar{t}}^6)$ system, which can be named ``dodecatoplet'', by analogy which the late pentaquark.
\section{Upper  bounds}
\subsection{Scaling}
It is well-known that under the $\mu\vec{r}_i\to \vec{r}_i$ transformation for all positions, and thus for the interquark distances $\vec{r}_{ij}$, the level energies of $H_N$, and in particular its ground-state, scale as
\begin{equation}\label{eq:sca}
E_N(m_t,\alpha_H, \mu)=\frac{\mu}{m_t}\,\epsilon_N(G)~, \qquad G=\frac{m_t \alpha_H}{\mu}~.
\end{equation}
Thus one is dealing with a one-parameter problem. For $\alpha_H=g_t^2/(4\pi)$, $m_t=172.6\;$GeV,  a reasonable Higgs mass  near $\mu=140\;$GeV, one gets $G\sim0.1$ as the order of magnitude of the dimensionless coupling.
\subsection{The two-body case}
\begin{figure}
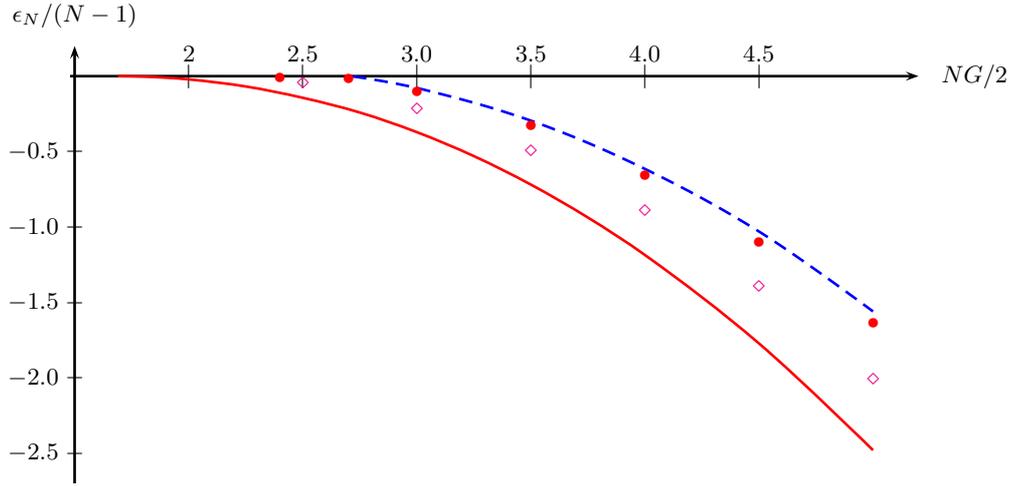

\begin{center}
\psset{xunit=3cm,yunit=2cm,linewidth=4pt}
\pspicture(1.45,-2.7)(5.5,.5)%\psgrid
\psline[arrows=->,linewidth=1pt](1.48,0)(5.2,0)\rput[l](5.3,0){\makebox{$NG/2$}}
%\psline[arrows=->,linewidth=1pt](0,-2.7)(0,.2)\rput[bc](0,.4){${\displaystyle\frac{\epsilon_N}{N-1}}$}
\psline[arrows=->,linewidth=1pt](1.5,-2.7)(1.5,.2)\rput[bc](1.5,.4){$\epsilon_N/(N-1)$}
\multido{\n=-2.5+.5}{5}{%
\rput[c](1.5,\n){\small$-$}
\rput[r](1.43,\n){$\n$}
}
\multido{\n=2+0.5}{6}{%
\rput[c](\n,0){\small$|$}
\rput[b](\n,.1){$\n$}
}
\pscurve[linecolor=red,linewidth=1pt](1.69,-.000146)(1.70,-0.000247)
(1.8,-0.003165)(1.9,-0.0101)(2,-0.02135)(2.25,-0.06841)(2.5,-.14345)
(2.75,-0.24114)(3,-0.37119)(3.25,-0.530)(3.5,-0.719)(3.75,-0.9373)(4.0,-1.1856)(4.5,-1.772)(5,-2.4799)
\pscurve[linecolor=blue,linewidth=1pt,linestyle=dashed]%
(2.72,-0.002)(2.74,-0.00629) (2.79,-.0178)(2.83,-0.0278)(2.89,-0.044)%
(2.95,-0.062)(3.,-0.0779)(3.5,-.295)(4,-.615)(4.5,-1.03)(5,-1.56)
\rput[c](5,-2.006){\magenta{$\diamond$}}
\rput[c](4.5,-1.392){\magenta{$\diamond$}}
\rput[c](4,-.889){\magenta{$\diamond$}}
\rput[c](3.5,-.498){\magenta{$\diamond$}}
\rput[c](3,-.215){\magenta{$\diamond$}}
\rput[c](2.5,-0.042){\magenta{$\diamond$}}

\rput[c](2.4,-0.015){\red{$\bullet$}}
\rput[c](2.7,-0.019){\red{$\bullet$}}
\rput[c](3,-0.102){\red{$\bullet$}}
\rput[c](3.5,-.330){\red{$\bullet$}}
\rput[c](4,-0.663){\red{$\bullet$}}
\rput[c](4.5,-1.0997){\red{$\bullet$}}
\rput[c](5,-1.64){\red{$\bullet$}}
\endpspicture
\end{center}
\caption{\label{fig:e2-12}  $\epsilon_N/(N-1)$, where $\epsilon_N$ is the $N$-body energy problem as a function of $G N/2$, where $G$ is the coupling. Solid curve: exact 2-body energy $\epsilon_2$ which is also a lower bound to $\epsilon_N/(N-1)$ for $N>2$. Dashed curve: Gaussian  variational approximation $\tilde\epsilon_N/(N-1)$, valid for any $N$.
{\magenta{$\diamond$}}:variational hyperscalar approximation $\bar{\epsilon}_{3}/2$ for the 3-body case. {\red{$\bullet$}}: hyperscalar approximation $\bar{\epsilon}_{12}/11$ for the 12-body case. }
\end{figure}

For two bosons, in a Yukawa potential, after scaling the variables and removing the centre of mass, the problem is reduced to 
\begin{equation}\label{eq:tb}
h_2=-\Delta -G \exp(-r)/r~,
\end{equation}
which starts supporting a bound state for $G\ge G_2\simeq 1.68$, as shown in the classic paper by Blatt and Jackson \cite{PhysRev.76.18}. 
With $G\sim 0.1$ we are thus far from a Yukawa binding of $(t\bar{t})$. 
The ground-state energy $\epsilon_2$  is drawn in Fig.~\ref{fig:e2-12} as a function of $G$.  
Also shown  is the variational upper bound $\tilde\epsilon_2=\min_a[t(a)-G\,p(a)]$ corresponding to a single normalised Gaussian, $\psi_a(\vec{r})\propto \exp(- a r^2/2)$, whose range parameter $a$ is optimised. The relevant expectation values, involving the complementary error function,  are
\begin{equation}\label{eq:1G}
%\begin{split}
t(a)=
%&=
\langle -\Delta\rangle=\frac{3a}{2}~,\qquad%\\
p(a)
%&=
=\langle v(r)\rangle
=
\frac{2\sqrt{a}}{\sqrt{\pi}}-
\exp\left[\frac{1}{4a}\right] \mathop{\rm erfc}\nolimits\left[\frac{1}{2 \sqrt{a}}\right]~.
%\end{split}
\end{equation}
This  trial function demonstrates binding for $G\gtrsim 2.71$ only, to be compared to the exact $G>G_2\simeq 1.68$.  This gives an idea of the validity of a simple variational approximation in a regime of weak binding. A function with better  behaviour at large $r$ would of course do much better.
\subsection{\boldmath Simple variational upper bound for the $N$-body system \unboldmath}
If one describes the relative motion using $(N-1)$ Jacobi variables 
\begin{equation}\label{eq:jac}
\vec{x}_1=\vec{r}_2-\vec{r_1}~,
\quad
\vec{x}_2=\frac{2 \vec{r}_3-\vec{r}_1-\vec{r}_2}{\sqrt{3}},\cdots
\end{equation}
from the rescaled positions, and removes the centre-of-mass motion, the $N$-body Hamiltonian reduces to 
\begin{equation}\label{eq:Nb}
-\sum_{i=1}^{N-1} \Delta_i -G\sum_{i<j} v(r_{ij})~.
\end{equation}
If the above Gaussian is generalised as $\prod_{i=1}^{N-1} \Psi_a(\vec{x}_i)$ and taken as trial function, 
the variational energy reads
\begin{equation}\label{eq:NbV1}
\tilde\epsilon_N=\min_a\left[ (N-1) t(a) -N(N-1) G p(a)/2\right]~,
\end{equation}
which is easily estimated, and becomes a better and better approximation as $N$ increases.
Note in (\ref{eq:NbV1}) the obvious relation 
\begin{equation}\label{eq:NbV2}
\tilde\epsilon_N(G)=(N-1)\,\tilde\epsilon_2(NG/2)~,
\end{equation}
which, as seen below, becomes a constraining inequality if the approximate energies are replaced by the  exact ones.
It also indicates that if the number $N$ of bosons is large enough, binding is achieved however small is the coupling $G$.
This possibility of binding systems whose subsystems are unbound, first pointed out by Thomas \cite{PhysRev.47.903}, is nowadays refereed to as ``Borromean binding'' . For references, see, e.g., \cite{Richard:2003nn}.

Note also that the variational energies obey the same scaling laws, virial theorem, etc., as the exact ones. This is stressed in~\cite{Kuchiev:2008fd}, and can be traced back up to very early papers dealing with quantum systems \cite{Hyl29,Foc30}.
\subsection{\boldmath Hyperscalar upper bound for $N$-body systems \unboldmath}
The above trial wave function $\prod_i \Psi_a(\vec{x}_i)$ is a particular function, namely Gaussian, of the hyperradius $r>0$ defined as
\begin{equation}\label{eq:NbV3}
r^2=\vec{x}_1^2+\cdots \vec{x}^2_{N-1}~.
\end{equation}
It occurs when the relative distances $\vec{x}_i$ are considered as a unique vector in a $3(N-1)$ dimensional space and
 polar coordinates $\{r, \Omega\}$ are introduced there to solve the Schr\"odinger equation. See, e.g.,  \cite{delaRipelle:2004ms}. As the potential $\sum v(r_{ij})$ is usually non central in this space, the Schr\"odinger equation in these coordinates is expressed as an infinite set of coupled radial equations. However, for a system of  bosons, the ground-state is well described in the (variational!) approximation consisting of retaining the lowest partial wave: this is the \emph{hyperscalar} approximation, which reads
\begin{equation}\label{eq:hypers}
u''(r)+\left[\bar{\epsilon}-\frac{L(L+1)}{r^2}+G\frac{N(N-1)}{2} v_{00}(r)\right]u(r)= 0~,
\end{equation}
with $\bar{\epsilon}\le \tilde\epsilon$ since the Gaussian approximation is a particular ansatz for $u(r)$.  Here, $L=3(N-2)/2$ is an effective orbital momentum, and the hypercentral projection reads
\begin{equation}\label{eq:hyperp}
v_{00}(r)=\frac{\displaystyle\int_0^{\pi/2} \cos^2\theta\,\sin^n\theta\, v(r\cos\theta)\,\mathrm{d}\theta}
{\displaystyle\int_0^{\pi/2} \cos^2\theta\,\sin^n\theta \,\mathrm{d}\theta}~,
\end{equation}
where $n=3N-7$. There is no difficulty to evaluate $v_{00}$ analytically%
\footnote{$\int v_{00}(r) r^{2L+2} \exp(-\alpha r^2)\,\mathrm{d} r/
\int  r^{2L+2} \exp(-\alpha r^2)\,\mathrm{d} r=p(\alpha)$ provides a good cross-check.}%
, but for large $N$, the analytic expression contains cancelling large positive and large negative terms, and a direct numerical estimate of (\ref{eq:hyperp}) might lead to better accuracy and stability in the computation.
\subsection{Results}
For (unphysical) $G=4$ and 
$N=12$, one gets $\tilde\epsilon\simeq -1096$ vs.\ $\bar\epsilon\simeq-1117$, which translates into $\langle H\rangle/[N(N-1) \alpha_H^2 m) \simeq-0.0480$ for $\mu /[(N-1) m \alpha_H]=0.022$ in the notation of the authors of Ref.~\cite{Kuchiev:2008fd}, very much compatible with their estimate based of a variational function $\prod\exp(-a r_i)$ and the result of a Hartree-Fock calculation. The  hyperscalar approximation is slightly below, i.e., better: the quality of the trial wave-functions in \cite{Kuchiev:2008fd} is partly compensated by the lack of removal of the centre-of-mass motion.

The variational energies $\tilde\epsilon(G)$ and $\bar\epsilon(G)$ in the physical domain (small $G$) are shown in Fig.~\ref{fig:e2:12}.
 A system of 12 bosons starts supporting a bound state for $G\gtrsim 0. 45$  from the Gaussian approximation  and the hyperspherical equation. The hyperscalar approximation is very close to the single-Gaussian approximation for $N=12$, while for the $N=3$ case, also shown in Fig.~\ref{fig:e2:12}, it is appreciably better.   The estimate of  \cite{Kuchiev:2008fd}, if translated in our notation (this is the $\mu_H\lesssim0.19$ of their figure, corresponds to   $G\gtrsim 0.478$. Note that Pacheco et al.~\cite{PhysRevA.39.4207,Epele} considered previously  ``self-Yukawian'' boson systems, using a self-consistent Hartree method similar to that of  \cite{Kuchiev:2008fd} and  obtained binding for 
a slightly better $G\gtrsim 0.438$. For $\alpha_H=1/(4\pi)$, this corresponds to a maximal Higgs mass of about $31\;$GeV.
\section{Lower bound for the ground-state energy}
Even though the above hyperspherical approximation and the central-field method of Refs.~\cite{Kuchiev:2008fd,PhysRevA.39.4207} are known to be very good approximations to the exact energy, the doubt could remain that the existence of bound states has been missed, or that the absolute value of the binding energy has been underestimated.  It is then desirable to derive a lower bound on the ground-state energy.

A basic tool consists of splitting the Hamiltonian into pieces, say
\begin{equation}\label{basic-split}
H=A+B+\cdots\ \Rightarrow\  E(H)\ge E(A)+E(B)+\cdots~,
\end{equation}
in an obvious notation where $E(H)$ is the ground-state energy of $H$. 
Saturation is obtained if $A$, $B$, etc., reach their minimum simultaneously.

The simplest application to our problem is done by rewriting the $N$-body Hamiltonian of (\ref{eq:H}) as
a sum of 2-body ones
\begin{equation} \label{eq:first-decomp}
H_N[m_t,\alpha_H]=\frac{1}{N-1}\,\sum_{i<j} H_2^{i,j}[m_t,(N-1)\alpha_H]~,
\end{equation}
and leads to
\begin{equation} \label{eq:first-ina}
E_N[m_t,\alpha_H]\ge \frac{N}{2}E_2[m_t, (N-1)\alpha_H]~,
\end{equation}
or in the rescaled variables
\begin{equation} \label{eq:first-inb}
\epsilon_N(G)\ge
\frac{N}{2} \epsilon_2[(N-1)G]~,
\end{equation}
but this is not very accurate because the energy of each two-body subsystem is bounded by its \emph{rest} energy, although each pair has an overall motion within the whole $N$-body system.
The remedy \cite{Post, Basdevant:1989pt,Epele} consists of writing identities among the intrinsic Hamiltonians
\begin{equation} \label{eq:Ht}
\widetilde H_N=H_N-\frac{(\vec{p}_1+\cdots+\vec{p}_N)^2}{2 N m}~,
\end{equation}
namely 
\begin{equation}
\label{decom-2}
\widetilde{H}_N(m_t,\alpha_H)=\frac{2}{N}\,\sum_{i<j}\widetilde{H}_2^{i,j}\left(m_t,N\alpha_H/2\right)~.
\end{equation}
This leads to the improved inequality
\begin{equation}\label{HP2}
E_{N}(m,\alpha_H)\ge (N-1)\,E_{2}\left(m,N\alpha_H/2\right)~,
\end{equation}
which is always better \cite{Basdevant:1989pt} than the previous inequality (\ref{eq:first-ina}).
 In the rescaled variables, it reads
\begin{equation}
\epsilon_N(G)\ge \underline{\epsilon}_N(G)=(N-1) \epsilon_2(N G/2)~.
\end{equation}

For the above numerical example with $G=4$, we obtain $ \underline{\epsilon}_{12}\simeq -1335$, to be compared with the upper bound $\bar\epsilon_{12}\simeq -1117$.  As seen in Fig.~\ref{fig:e2-12}, the window is rather narrow between the variational upper bound and the Hall--Post lower bound.

But we are here in a regime of deep binding that would require drastic relativistic corrections.
A more important consequence  of (\ref{decom-2}) is that $H_N$ hardly supports bound states if each $H_2$ is positive. This means that binding requires $G> G_N$ with 
\begin{equation}
\label{eq:cond-G}
G_N\ge \frac{ 2}{N}\,G_2\simeq 0.28~,
\end{equation}
which corresponds (again for $\alpha_H=1/(4\pi)$ to a minimal Higgs mass which is certainly below 49\,GeV.
\section{Summary}
We have revisited the problem of binding 6 top quarks and 6 top antiquarks with  a Yukawa potential mediated by Higgs exchange. The problem depends only of the dimensionless coupling $G=m_t\alpha_H/\mu$. By combining variational estimates and Hall-Post type of lower bound, the 12-body problem is under control. In particular, it is found that the minimal strength to bind the system belongs to the interval
\begin{equation}
0.28\le G_{12}^\text{min}\le 0.43~.
\end{equation}
With a Higgs coupling taken as $g_t=1$, corresponding to $\alpha_H=1/(4\pi)$, the maximal Higgs mass leading to this binding should be 
\begin{equation}
31\,\text{GeV} < \mu^\text{max} <  49\,\text{GeV}~.
\end{equation}
With the current ideas on the Higgs mass and its coupling, the existence of such dodecatoplet seems very unlikely. Perhaps higher systems can be bound. For $^3\mathrm{He}$ atoms, the interaction potential contains attractive parts, but is unable to bind the dimer, unlike the $^4\mathrm{He}$ case, where the dimer is bound, due to the heavier mass of the constituents. It was found, however, that binding becomes possible for $N\gtrsim 35$ atoms of $^3\mathrm{He}$ \cite{PhysRevLett.84.1144,Bressanini}. Of course, to study $(t^n\bar{t}^m)$ systems with $n$ or $m$ larger than $6$, one should account for the antisymmetrisation effects in building the orbital wave function.
%\bibliographystyle{h-physrev4}
%\bibliography{Dodeca}

\end{document}